# Effect of low-intensity pulsed ultrasound on biocompatibility and cellular uptake of chitosan-TPP nanoparticles

Junyi Wu

Stony Brook University, Department of Materials Science and Engineering, Stony Brook, New York 11794

Gaojun Liu

Stony Brook University, Department of Materials Science and Engineering, Stony Brook, New York 11794

Yi-Xian Qin

Stony Brook University, Department of Biomedical Engineering, Stony Brook, New York 11794

Yizhi Meng[a]

Stony Brook University, Department of Materials Science and Engineering, 314 Old Engineering, Stony Brook, New York 11794

[a]Electronic mail: yizhi.meng@stonybrook.edu

Using low molecular weight chitosan nanoparticles (CNPs) prepared by an ionic gelation method, we report the effect of low-intensity pulsed ultrasound (US) on cell viability and nanoparticle uptake in cultured murine pre-osteoblasts. Particle size and zeta potential are measured using Dynamic Light Scattering (DLS), and cell viability is evaluated using the MTS assay. Results show that 30 min delivery of CNPs at 0.5 mg/mL is able to prevent loss of cell viability due to either serum starvation or subsequent exposure to US (1 W/cm$^2$ or 2 W/cm$^2$, up to 1 min). Additionally, flow cytometry data suggest that there is a close association between cellular membrane integrity and the presence of CNPs when US at 2 W/cm$^2$ is administered.

## I. INTRODUCTION

A linear polysaccharide derived from the exoskeleton of crustaceans, cell walls of fungi and cocoons of insects,[1] chitosan is a low-immunogenic cationic polymer that can form solid colloidal nanoparticles in the range of 1 nm to 1000 nm.[2] Chitosan-based nanoparticles (CNPs) are highly biocompatible [3-6] and respond to external stimuli such as ultrasound (US).[7,8] Because US irradiation is widely used in clinical diagnostics as well as for wound healing and cancer therapy,[9-11] polymers that are capable of undergoing sonolytic degradation are of great biomedical importance.

US is composed of a propagating pressure wave or sound wave, which can help transfer mechanical energy into various tissues of the body.[12] The frequency, duty cycle and duration all contribute toward the total absorbance of energy. US has been found to enhance bone fracture healing[13] as well as aid in drug release.[14] Low-intensity US in particular can enhance the delivery efficiency of drug carriers such as polyethylene glycol (PEG), liposomes and micelles, thereby increasing the therapeutic efficacy of the cargo.[15-18] For example, US (0.3 W/cm$^2$) has been shown to significantly increase the anticancer effect of doxorubicin (DOX) in human bladder carcinoma cells.[19] Exposure to US at 1 MHz (0.2-0.5 W/cm$^2$) for 60 s was also shown to increase sensitivity to doxorubicin in DOX-resistant human uterine sarcoma cells.[20] As such, US-mediated delivery of nanotherapeutics may help reverse multidrug-resistance of certain cancer cells.[21,22]

Although low-intensity US has been shown to be effective in assisting drug delivery from polymer-based vehicles, there is little documentation regarding the viability of the healthy tissue at the irradiation site. The primary objective of this work therefore is to examine the effect of ultrasound-assisted delivery of chitosan nanoparticles (CNPs) in a mammalian cell line by investigating cell viability and uptake. A more complete understanding of non-invasive sonodynamic therapy can help evaluate its efficacy as a potential adjuvant to conventional drug delivery.

## II. EXPERIMENTAL

### A. Materials

Low molecular weight chitosan (deacetylation degree 75%-85%), sodium tripolyphosphate (TPP, technical grade, 85%), 2% ninhydrin reagent solution (ninhydrin and hydrindantin in DMSO and lithium acetate buffer, pH=5.2), and fluorescein 5(6)-isothiocyanate were purchased from Sigma-Aldrich (St. Louis, MO). Sodium hydroxide pellets (reagent grade, ≈98.8%), formaldehyde solution (Reagent A.C.S, ≈37%) and glacial acetic acid (analytical grade, ≈99.9%) were obtained from J.T. Baker (Center Valley, PA). Methanol (Reagent A.C.S, ≈99.9%) and water (HPLC grade) were purchased from Pharmco-AAPER (Brookfield, CT). D-(+)-Trehalose dihydrate (≈99%) was obtained from Acros Organics (NJ, USA) and dialysis cassettes (3500 MWCO) were purchased from Thermo Scientific (Rockford, IL).

## B. Synthesis

### 1. Chitosan-TPP nanoparticles

An ionic gelation method was used to synthesize the chitosan nanoparticles, similar to previous reports.[2,23] Low molecular weight chitosan was dissolved in acetic acid (1% v/v, in water) for 30 min under constant stirring to make a 0.5% (w/v) chitosan stock solution, which was filtrated with a 0.8 µm syringe filter before gelation. A few drops of sodium hydroxide were added to the chitosan stock solution to adjust the pH value to 4.6-4.8. Tripolyphosphate (TPP) solution (0.25% w/v, in water) was added dropwise into the chitosan solution until opalescent droplets could be seen. After 5 min of reaction, the nanoparticle solution was centrifuged at 2700 RPM for 30 min and rinsed extensively with DI water. D-(+)-Trehalose was added to nanoparticle solution to a final concentration of 5% (w/v). The nanoparticles were stored in a -80°C freezer for at least 24 h and lyophilized. To prepare FITC-labeled CNP (FITC-CNP), FITC-conjugated chitosan was first prepared using published protocols.[24] The same ionic gelation method was conducted in the dark.

## C. Characterization
### 1. FITC labeling efficiency

In order to calculate the labeling efficiency of FITC, a standard curve plotting absorbance vs. FITC concentration was used. FITC standard solutions ranging from 0.01 µg/mL to 0.08 µg/mL were prepared by diluting 100 µg/mL FITC stock solutions (in methanol) with 1x phosphate-buffered saline (PBS). An FITC-chitosan test sample was prepared by dissolving FITC-chitosan in 0.1 M acetic acid followed by dilution with PBS, until a final concentration of 1.0 µg/mL was achieved. All solutions were evaluated in triplicate by ultraviolet-visible spectroscopy (Perkin Elmer Lambda 950, Perkin Elmer, Waltham, MA). FITC labeling efficiency was calculated using the equation below:

$$\text{FITC labeling efficiency} = \frac{\text{weight of FITC}}{\text{weight of FITC - weight of chitosan}} \times 100\% \quad (1)$$

### 2. Particle size and zeta potential

Measurements of particle size and surface charge (characterized as zeta potential) were performed by dynamic light scattering (DLS), using a ZEN3600 Nano Zetasizer (Malvern Instruments, UK) located at the Center for Functional Nanomaterials (CFN) of Brookhaven National Laboratory (Upton, NY). Freeze-dried chitosan nanoparticles prepared at 6:1 chitosan:TPP mass ratio were dissolved in DI water to make a 1 mg/mL CNP suspension. In order to remove large aggregates, the suspension was then passed through a 0.8 µm syringe filter, and the filtrate was collected and stored at 4°C prior to analysis. The particle size measurements were conducted at room temperature and each measurement lasted for 120 s. For zeta potential measurement, a specific zeta dip cell was used and 30 measurements were collected for each sample. All measurements were performed at 25°C in triplicate.

### D. Nanoparticle delivery and ultrasound irradiation

MC3T3-E1 murine pre-osteoblasts (subclone 4; American Type Culture Collection, Manassas, VA) were maintained in α-modified minimum essential medium (α-MEM; Gibco, Grand Island, NY) supplemented with 10% fetal bovine serum (FBS) and 1% penicillin-streptomycin (pen-strep) solution. The cells were maintained at 37°C in an atmosphere of 5% carbon dioxide and 95% relative humidity.

Twenty-four hours prior to the delivery of chitosan-TPP nanoparticles, MC3T3-E1 cells were seeded onto a 96-well tissue culture plate at 10,000 cells/cm$^2$ (for viability assay), a 24-well tissue culture plate at 5,000 cells/cm$^2$ (for confocal microscopy), or a 35 mm (diam.) tissue culture plate at 50,000 cells/cm$^2$ (for flow cytometry). Cells were incubated for 24 h in a humidified incubator (5% $CO_2$) to attach completely. The supernatant was aspirated and the cells were washed with PBS. Fresh complete medium (supplemented with FBS) or serum-free medium containing either 0 or 0.5 mg/mL CNPs was added. The plates were gently swirled before being returned to the incubator. After 30 min, cells that were to receive ultrasound (US) treatment were removed from the incubator and immediately sonicated by a Sonicator 740 (Mettler Electronic Corp. Anaheim, CA) with a 1 MHz applicator and 5 cm$^2$ probe (Fig. 1). US intensity was set at either 1 W/cm$^2$ or 2 W/cm$^2$, for a total exposure time of either 30 s or 60 s. Cells that were not treated with US were kept on the lab bench for the same amount of time. All cells were then returned to the incubator for an additional 90 minutes.

## *E. Cell viability*

Cytotoxic effect of ultrasound and CNP was examined using the MTS viability assay. Aliquots of MTS/PMS solution were added to the supernatant in each cell culture well, and the tissue culture plate was swirled gently before being incubated for 1 h at 37°C. Absorbance was immediately read at 490 nm using a microplate reader (BioTek FLx800; BioTek, Winooski, VT).

## F. *Confocal microscopy*

Cells were fixed with 3.7% formaldehyde (in PBS) for 15 min, followed by extensive rinsing with PBS and immersion in 2.5 µg/mL DAPI solution for 5 min. After removal of the DAPI solution, the cells were rinsed two more time with PBS. Fluorescence micrographs were then captured using a DSU confocal microscope (Olympus IX2-DSU) and Z-scans were obtained with a total scan depth of 5.8 µm, with a 0.2 µm step size.

## G. *Flow cytometry*

Cells that were exposed to 2 W/cm$^2$ US for 60 s were detached from the tissue culture plate with 0.05% Trypsin-EDTA and resuspended in 1 mL complete medium. Propidium iodide (10 µg/mL in PBS) was applied for 1 min at 4°C in the dark. Samples were immediately read by a FACS Calibur flow cytometer (BD Biosciences) using a 488 nm laser.

# III. RESULTS AND DISCUSSION

## A. *Physicochemical Properties of Chitosan-TPP Nanoparticles*

The FITC labeling efficiency of FITC-chitosan calculated using Equation (1) was 2.4%, which is similar to previous reported value of 2.7%.[25] Average particle size of freeze dried CNPs and FITC-CNPs re-suspended in water was 288.4 ± 1.2 nm and 294.0

± 10.8 nm, respectively. Average zeta potential of CNP and FITC-CNP in water was 36.87 0 .27 mV and 32.90 ± 0.64 mV, respectively.

## B.  In Vitro Uptake of FITC-labeled Chitosan-TPP Nanoparticles

Confocal micrographs showed greater internalization of the FITC-CNP by the MC3T3-E1 cells under US exposure compared to the non-irradiated cells (Fig. 2). Aggregates of FITC-CNP appeared larger in cells that were treated for either 30 seconds at 1 W/cm$^2$ or 60 s at 2 W/cm$^2$.

## C.  Cell Viability

Results from the MTS viability assay showed that when the untreated control MC3T3-E1 cells were cultured in complete medium ("CM" group, Fig. 3) without exposure to either US or CNP, cell density after 2 h of attachment reached the seeding density of 10,000 cells/cm$^2$ (Fig. 3), indicating 100% viability. Application of US at either 1 or 2 W/cm$^2$ for a minimum of 30 s to cells maintained in complete medium significantly decreased cell density by ~20% ($p < 0.001$; Fig. 3). There was no significant difference among the US treatments. When cells were temporarily serum-starved but not exposed to CNP ("SS" group, Fig. 3), cell density at 2 h dropped to 6600-6800 cells/cm$^2$ for all samples irrespective of US treatment, corresponding to a 32-34% reduction relative to untreated controls ($p < 0.001$). Interestingly, when CNPs were added during the 2 h serum starvation period ("SS+CNP" group, Fig. 3), cells appeared to be more viable compared to those that were serum-starved for the same duration in the absence of CNP. More specifically, area densities of cells were not significantly different between the untreated controls and those exposed simultaneously to CNP and either lower intensity US (1 W/cm$^2$) or no US at all (p > 0.2; Fig. 3). At the higher US intensity of 2 W/cm$^2$, cell density decreased significantly by 8% after 30 s of exposure (p < 0.001) and 13% after 60 s of exposure (p < 0.01).

## D. Flow Cytometry

To further investigate the significant decrease in cell viability after the MC3T3-E1 cells were exposed to 60 s of US at 2 W/cm$^2$, we conducted flow cytometry measurements following the addition of propidium iodide (PI) to the treated cells. Scatter plots indicated that US alone did not have an effect on the cells' ability to uptake PI (Fig. 4 a-b and e-f), which suggests that cellular membrane was not damaged by US, even though metabolic activity was decreased as seen from the MTS data (Fig. 3). When CNPs were delivered alone in the absence of US stimulation, FITC fluorescence was notably stronger, as shown by the dense scatter in the lower-right quadrant (Fig. 4 c and g). This suggests that CNPs had a great affinity for the MC3T3-E1 cells. We also observed that uptake of PI was diminished in this case, as seen by an absence of dense clusters in the upper-left quadrant of the scatter plot (Fig. 4 c and g) compared to either untreated cells (Fig. 4 a and e) or those exposed to US alone (Fig. 4 b and f). This indicates that CNPs may have protected the cell membrane from being damaged during sample preparation. When US was applied immediately after CNP delivery, FITC fluorescence was even more greatly enhanced (lower-right quadrant, Fig. 4 d and h), indicating that more CNPs were associated with, and possibly internalized by, the MC3T3-E1 cells. In addition, PI fluorescence was also enhanced (upper-right quadrant, Fig. 4 d and h), which suggests that cellular membrane may have become less intact. Moreover, because the population of cells with strong PI fluorescence overlapped those with strong FITC fluorescence (Fig.

4 d and h), we believe that membrane leakiness may be associated with the presence of CNPs.

## IV. SUMMARY AND CONCLUSIONS

In this study, chitosan/TPP (CNP) and FITC-labeled chitosan/TPP nanoparticles (FITC-CNP) were successfully formulated by a modified ionic gelation method. FITC-CNPs were delivered to murine pre-osteoblasts (MC3T3-E1 cell line, subclone 4), and the uptake was shown to be enhanced by ultrasound (US) irradiation. Serum starvation for 2 h resulted in a 32-34% reduction in cell viability, which was not observed when CNPs were administered simultaneously. The same phenomenon was observed when low-intensity US treatment (1 W/cm$^2$) was applied during serum starvation, but at higher intensity (2 W/cm$^2$) cell viability was notably decreased. Additionally, there seems to be a close association between ultrasound-assisted CNP delivery and membrane integrity. Taken together, these results suggest that CNPs may alter the sensitivity of pre-osteoblasts to therapeutic levels of low-intensity pulsed ultrasound.

## ACKNOWLEDGMENTS

Research was carried out in part at the Center for Functional Nanomaterials, Brookhaven National Laboratory, which is supported by the U.S. Department of Energy, Office of Basic Energy Sciences, under Contract No. DE-AC02-98CH10886.

**Figure Captions**

Figure 1. Schematic of experimental setup.

Figure 2. Fluorescence micrographs showing FITC-labeled chitosan nanoparticles (visualized in the green channel) after being internalized by MC3T3-E1 preosteoblasts (nuclei visualized in the blue channel). Cells were exposed to nanoparticles in the absence (a,b) or presence (c-f) of ultrasound and incubated for a total of 30 minutes (a) or 120 minutes (b-f). Ultrasound was applied at 1 W/cm$^2$ for 30 s (c) or 60 s (d), and at 2 W/cm$^2$ for 30 s (e) or 60 s (f). Images are representative of two independent experiments. Scale, 25 µm.

Figure 3. Effect of ultrasound on viability of MC3T3-E1 cells 24 hours after CNP delivery and US stimulation. All values were expressed as mean±SE, **p<0.01, ***p<0.001 (n=3).

Figure 4. Scatter (a-d) and contour (e-h) plots of flow cytometry data for untreated cells (a,e), cells exposed only to 2 W/cm$^2$ US for 60 s (b,f), cells exposed only to chitosan nanoparticles (c,g), and cells exposed both to chitosan nanoparticles and 2 W/cm$^2$ US for 60 s (d,h). Fluorescence intensities of propidium iodide (PI) and FITC are displayed on the x- and y-axes, respectively.

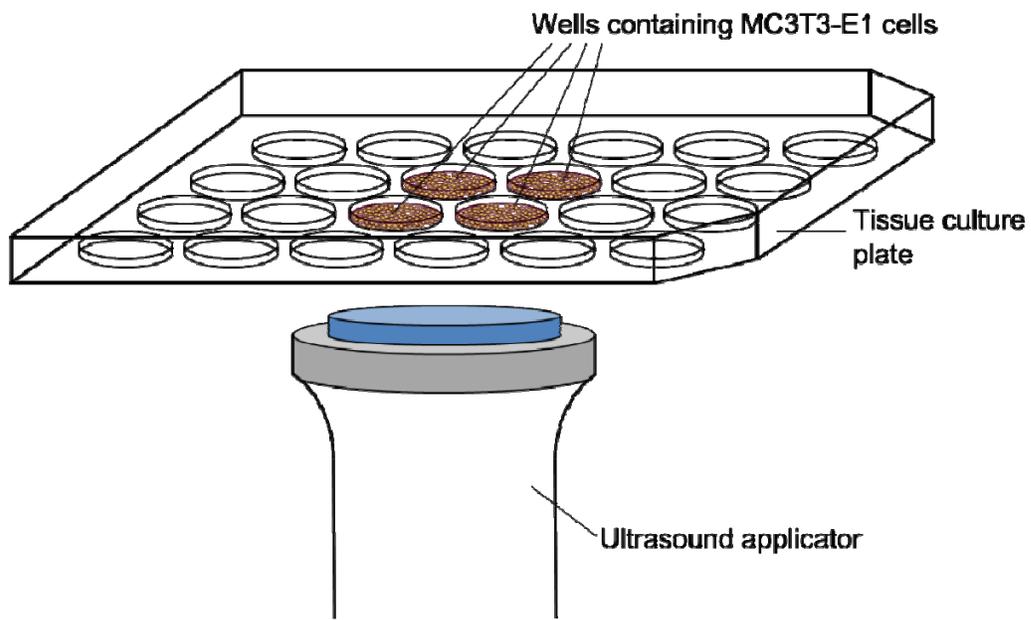

Figure 1

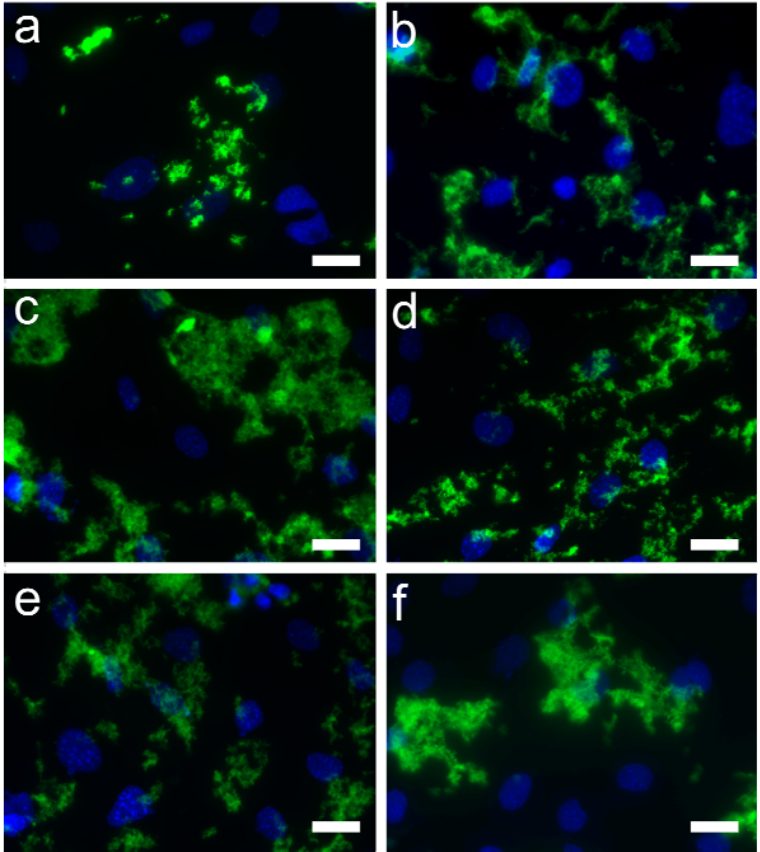

Figure 2

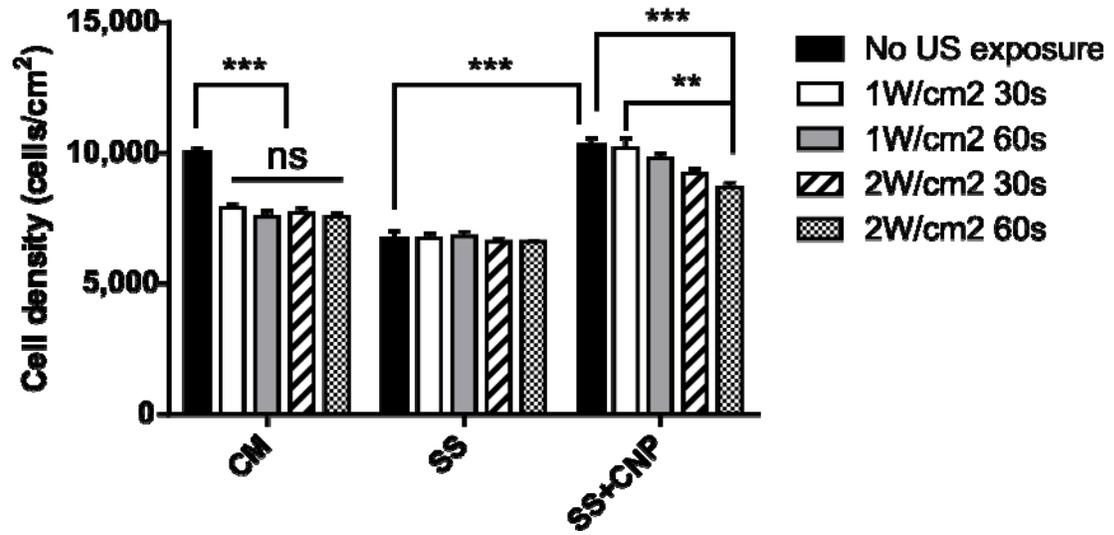

Figure 3.

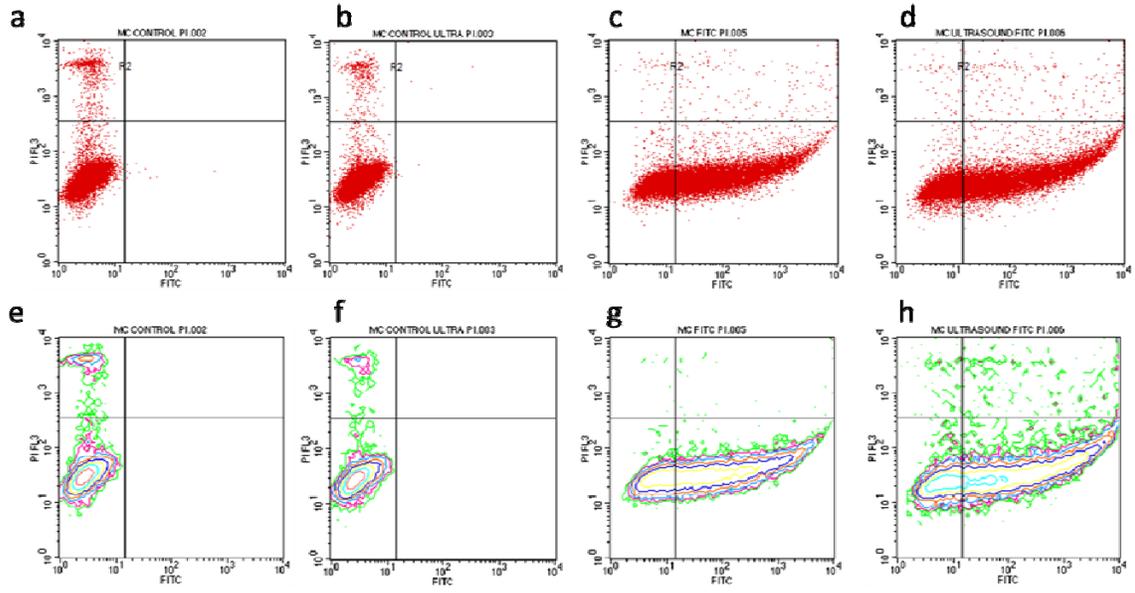

Figure 4